\documentclass[journal]{IEEEtran}
\usepackage{stfloats}
\usepackage{cite}
\usepackage{float}
\usepackage{graphicx}
\usepackage{subfigure}
\usepackage{color}
\usepackage{amssymb}
\usepackage{amsmath}
\usepackage{algorithm}
\usepackage{algorithmic}
\usepackage{bm}
\begin{document}

\title{Joint Optimization for Achieving Covertness in MIMO Over-the-Air Computation Networks}

\author{Junteng Yao, Tuo Wu, Ming Jin, \emph{Member}, \emph{IEEE}, Cunhua Pan, \emph{Senior Member}, \emph{IEEE},  Quanzhong Li, and Jinhong Yuan, \emph{Fellow}, \emph{IEEE}

\thanks{\emph{(Corresponding author: Tuo Wu and Cunhua Pan.)}}

\thanks{J. Yao and M. Jin are with the Faculty of Electrical Engineering and Computer Science, Ningbo University, Ningbo 315211, China (e-mail: \{yaojunteng, jinming\}@nbu.edu.cn).}

\thanks{T. Wu is with the School of Electronic Engineering and Computer Science at Queen Mary University of London, London E1 4NS, U.K. (e-mail: tuo.wu@qmul.ac.uk).}

\thanks{C. Pan is with the National Mobile Communications Research Laboratory, Southeast University, Nanjing 210096, China. (e-mail: cpan@seu.edu.cn).}

\thanks{Q. Li is with the School of Computer Science and Engineering, Sun Yat-sen University, Guangzhou 510006, China (e-mail: liquanzh@mail.sysu.edu.cn).}

\thanks{J. Yuan is with the School of Electrical Engineering and Telecommunications,  University of New South Wales, Sydney, NSW 2052, Australia (e-mail: j.yuan@unsw.edu.au).}
}

\maketitle

\begin{abstract}
This paper investigates covert data transmission within a multiple-input multiple-output (MIMO) over-the-air computation (AirComp) network, where sensors transmit data to the access point (AP) while guaranteeing covertness to the warden (Willie). Simultaneously, the AP  introduces  artificial noise (AN) to confuse Willie, meeting the covert requirement. We address the challenge of minimizing mean-square-error (MSE) of the AP, while considering transmit power constraints at both the AP and the sensors, as well as ensuring the covert transmission to Willie with a low detection error probability (DEP). However, obtaining globally optimal solutions for the investigated non-convex problem is challenging due to the interdependence of optimization variables. To tackle this problem, we introduce an exact penalty algorithm and transform the optimization problem into a difference-of-convex (DC) form problem  to find a locally optimal solution. Simulation results showcase the superior performance in terms of our proposed scheme in comparison to the benchmark schemes.
\end{abstract}


\section{Introduction}
The advent of the sixth-generation (6G) networks has amplified the requirement for enhanced connectivity within Internet of Things (IoT) networks \cite{AGiridhar06,KWChoi18}. Utilizing the superposition property of wireless multiple access channels, over-the-air computation (AirComp) enables rapid data aggregation and seamlessly integrates communication and computation processes to facilitate extensive connectivity among IoT devices \cite{WLiu20,NYan,BNazer07}. In a single-input single-output (SISO) AirComp system, Cao \emph{et al.} \cite{XCao20} proposed to minimize the mean-square-error (MSE) by designing the signal scaling factor at the access point (AP) and the transmit power at sensors.
Li \emph{et al.} \cite{YLi22} minimized MSE of the AP in the multiple-input single-output (MISO) AirComp system, where both the direct link and relay link were considered. Moreover, intelligent reflecting surface (IRS)-assisted AirComp networks\cite{WFang21}, unmanned aerial vehicle (UAV)-assisted AirComp networks \cite{MFu22} were also studied.

The extensive data transmissions among IoT devices within AirComp networks pose a potential risk of information leakage due to the inherent openness of wireless propagation. In an attempt to mitigate this concern, Hu \emph{et al.} \cite{CHu22} studied physical layer security (PLS) in the AirComp networks, where the MSE of the eavesdropper is larger than a predefined threshold. While conventional PLS can partially impede the eavesdropper's effectiveness, it falls short of adequately addressing privacy concerns associated with data transmission in AirComp networks. For instance, some certain  scenarios, such as private or military transmissions, require to safeguard the presence of the transmission from detection or exposure \cite{BABash15}. Collaborative efforts are necessary to conceal the transmission's presence in AirComp networks when sensors transmit sensitive data to the AP.
Therefore, covert transmission has arisen as a promising technique for achieving covert operations within AirComp networks \cite{JHu19,SYan19,SMa21}. To the best of our knowledge, research on covert transmission within multiple-input multiple-output (MIMO) AirComp networks is currently lacking.

Motivated by the above discussion, this paper aims to investigate the minimization of MSE of the AP within an MIMO AirComp network under the covertness constraint. The main contributions of this paper are summarized as follows:
\begin{itemize}
\item We consider the covert transmission in MIMO AirComp networks, and formulate MSE minimization problem of the AP under the constraints of the transmit power at the AP and the sensors, as well as the detection error probability (DEP) requirement of a warden named Willie. Here, the AP employs artificial noise (AN) to enhance covertness against Willie.
\item By using the Kullback-Leibler divergence (KLD), we convert the DEP constraint into a suitable convex form. However, the formulated optimization problem is still non-convex due to the coupled variables. To address this issue, we propose the exact penalty  algorithm to obtain a locally optimal solution.
\item Numerical results demonstrate that our proposed scheme outperforms other benchmark schemes in terms of MSE performance.
\end{itemize}

\emph{Notations}: The conjugate transpose, trace, and Frobenius norm are denotes as $\mathbf{A}^H$, $\mathrm{tr}(\mathbf{A})$, and $\left\|\mathbf{A}\right\|$, respectively. $\mathbf{A}\succeq (\succ) \mathbf{0}$ represents that $\mathbf{A}$ is positive semidefinite (positive definite). $\mathbb{R}\{x\}$ means the real part of $x$. $\mathbb{E}\{\cdot\}$ denotes the expectation operation. $\mathcal{CN}(\mathbf{0}, \mathbf{I})$ denotes a random vector following the distribution of mean $\mathbf{0}$ and covariance $\mathbf{I}$. $\mathcal{CN}(0,\sigma^2)$ denotes the distribution of a circularly symmetric complex Gaussian random variable with mean 0 and variance $\sigma^2$.

\section{System Model and Problem Formulation}
\subsection{System Model}
We investigate an MIMO covert AirComp system comprised of $K$ sensors, a full-duplex (FD) AP, and a warden named Willie. Each sensor is equipped with $N_s$ antennas, while the AP is equipped with $N_r$  antennas for receiving sensor-transmitted signals and $N_t$ additional antennas for generating artificial noise (AN) to confuse Willie. Besides, Willie is equipped with one antenna.  The sensors seek to transmit messages to the AP covertly to evade detection by Willie, whose goal is to identify any transmissions between the sensors and the AP.

We denote the channels connecting the $k$th sensor to the AP, the $k$th sensor to Willie, the AP to Willie as $\mathbf{H}_k\in{\mathbb{C}^{N_r\times{N_s}}}$, $\mathbf{g}_k\in{\mathbb{C}^{1\times{N_s}}}$, and $\mathbf{f}\in{\mathbb{C}^{1\times{N_t}}}$, respectively. Moreover, the self-interference channel at the AP owing to the FD mode is represented as $\mathbf{H}_{aa}=\sqrt{\rho}\mathbf{H}$, where $\mathbf{H}\in{\mathbb{C}^{N_r\times{N_t}}}$ denotes the feedback loop link of the AP and $\rho$ represents the self-interference coefficient. We assume that all channels are quasi-static independent block-fading channels.

\subsection{MSE Analysis at the AP}
Let us denote the pre-processing signal of the $k$th sensor as $\mathbf{s}_k\in{\mathbb{C}^{N_s\times{1}}}$, $k\in\mathcal{K}=\{1, 2, \cdot\cdot\cdot, K\}$, which is assumed to be normalized and independent of each other, i.e., $\mathbb{E}(\mathbf{s}_k\mathbf{s}^H_k) = \mathbf{I}_{N_s}$ and $\mathbb{E}(\mathbf{s}_k\mathbf{s}^H_m) = \mathbf{0}, \forall k, m \in\mathcal{K}, k\neq m$. The AP adopts the summation  operation as the target-function, which is expressed as \cite{CHu22}
\begin{align}\label{q1}
\mathbf{s}=\sum_{k=1}^{K} \mathbf{s}_k.
\end{align}

We denote $\mathbf{W}_k\in{\mathbb{C}^{N_s\times{N_s}}}$, $\mathbf{V}\in{\mathbb{C}^{N_t\times{N_t}}}$ and $\mathbf{z}\in{\mathbb{C}^{N_t\times{1}}}$ as the transmit beamforming matrix at the $k$th sensor, the AN beamforming matrix and AN signal at the AP, respectively. Additionally, it is assumed that $\mathbb{E}(\mathbf{z}\mathbf{z}^H) = \mathbf{I}_{N_t}$. Therefore, the signal received at the AP is
\begin{align}\label{q3}
\mathbf{y}_a&=\sum_{k=1}^{K} \mathbf{H}_k\mathbf{W}_k\mathbf{s}_k+\mathbf{H}_{aa}\mathbf{V}\mathbf{z}+\mathbf{n}_a,
\end{align}
where $\mathbf{n}_a\sim \mathcal{CN}(\mathbf{0}, \sigma^2_a\mathbf{I}_{N_r})$ denotes the additive white
Gaussian noise (AWGN) at the AP.

By applying the aggregation beamforming, the computing output of the AP can be expressed as
\begin{align}\label{q5}
\hat{\mathbf{s}}&=\mathbf{U}^H_a\mathbf{y}_a,
\end{align}
where $\mathbf{U}_a\in{\mathbb{C}^{N_r\times{N_s}}}$ denotes the aggregation matrix at the AP.

To measure the computation distortions between $\mathbf{s}$ and $\hat{\mathbf{s}}$, we employ the MSE as a performance metric for the AirComp networks \cite{BNazer07,XCao20}. The MSE of the AP is defined as
\begin{align}\label{q7}
\mathrm{MSE}_a=&\mathbb{E}(||\mathbf{s}-\hat{\mathbf{s}}||^2)\nonumber\\
=&\sum_{k=1}^{K}||\mathbf{U}^H_a\mathbf{H}_k\mathbf{W}_k-\mathbf{I}_{N_s}||^2+||\mathbf{U}^H_a\mathbf{H}_{aa}\mathbf{V}||^2\nonumber\\
&+\sigma^2_a||\mathbf{U}^H_a||^2.
\end{align}

\subsection{Covertness Analysis at Willie}
Suppose $\mathcal{H}_1$ represents the hypothesis that the sensors have transmitted signals, while $\mathcal{H}_0$ represents the hypothesis that no AirComp has occurred. Consequently, the received signal at Willie can be expressed as
\begin{equation}\label{aq5}
y_w
=\left\{ \begin{array}{cc}
\mathbf{f}\mathbf{V}\mathbf{z}+n_w, &  \mathcal{H}_0,\\
\sum_{k=1}^{K} \mathbf{g}_k\mathbf{W}_k\mathbf{s}_k+\mathbf{f}\mathbf{V}\mathbf{z}+n_w, & \mathcal{H}_1,
\end{array}\right.
\end{equation}
where $n_w$ is the AWGN at Willie.

As Willie needs to make a binary decision between the two hypotheses in \eqref{aq5}, the distributions of $y_w$ are given by
\begin{equation}
y_w
\sim \left\{ \begin{array}{cc}
\mathcal{CN}(0, \sigma_0), & \mathcal{H}_0,\\
\mathcal{CN}(0, \sigma_1), & \mathcal{H}_1,
\end{array}\right.
\end{equation}
where
\begin{align}
\sigma_0&=\mathbf{f}\mathbf{V}\mathbf{V}^H\mathbf{f}^H+\sigma^2_w,\\
\sigma_1&=\sum_{k=1}^{K} \mathbf{g}_k\mathbf{W}_k\mathbf{W}^H_k\mathbf{g}^H_k+\mathbf{f}\mathbf{V}\mathbf{V}^H\mathbf{f}^H+\sigma^2_w.
\end{align}

Let us define $p_0(y_w)=f(y_w|\mathcal{H}_0)$ and $p_1(y_w)=f(y_w|\mathcal{H}_1)$ as the likelihood functions of $y_w$ under $\mathcal{H}_0$ and $\mathcal{H}_1$respectively, we have
\begin{align}\label{q9}
p_i(y_w)=\frac{1}{\pi\sigma_i}\exp(-|y_w|^2/\sigma_i), i\in\{0, 1\}.
\end{align}

We assume that the priori probabilities of hypotheses $\mathcal{H}_0$ and $\mathcal{H}_1$ are equal. The DEP, which is adopted to measure the detection performance of Willie, can be defined as
\begin{align}
\eta=\mathbb{P}(D_1|\mathcal{H}_0)+\mathbb{P}(D_0|\mathcal{H}_1),
\end{align}
where $D_0$ and $D_1$ denote the decisions made by Willie corresponding to the hypotheses $\mathcal{H}_0$ and $\mathcal{H}_1$, and $\mathbb{P}(D_1|\mathcal{H}_0)$  and $\mathbb{P}(D_1|\mathcal{H}_0)$ are the false alarm probability and the miss detection probability, respectively.
 
 By using Pinsker's inequality \cite{SYan19}, a lower bound of $\eta$ is given by
\begin{align}
\eta\geq 1-\sqrt{\frac{1}{2}\mathbb{D}(p_0||p_1)},
\end{align}
where $\mathbb{D}(p_0||p_1)$ denotes the KLD from $p_0$ to $p_1$. Moreover, the expression of $\mathbb{D}(p_0||p_1)$ is given by \cite{SMa21}
\begin{align}
\mathbb{D}(p_0||p_1)&=\ln\frac{\sigma_1}{\sigma_0}+\frac{\sigma_0}{\sigma_1}-1.
\end{align}

To ensure successful covert transmission with a predefined tolerated detection coefficient $\epsilon>0$, $\mathbb{D}(p_0||p_1)$ should satisfy the following condition \cite{JHu19}
\begin{align}
\mathbb{D}(p_0||p_1)\leq 2\epsilon^2.
\end{align}
\subsection{Problem Formulation}
In this paper, we consider the MIMO AirComp system under the covertness constraint. Therefore, the objective is to minimize the AP's MSE while accounting for the transmission power constraints at the AP and sensors, along with the DEP requirement of Willie. The problem is formulated as follows
\begin{subequations}\label{q10}
\begin{align}
\min_{\scriptsize\begin{array}{c}\mathbf{W}_k, \mathbf{V}, \mathbf{U}_a
 \end{array} }\ &\mathrm{MSE}_a\label{q10a}\\
\textrm{s.t.}\quad \quad \  & \mathbb{D}(p_0||p_1)\leq 2\epsilon^2,\label{q10b}\\
&||\mathbf{W}_k||^2\leq P_k, k\in \mathcal{K},\label{q10c}\\
&||\mathbf{V}||^2\leq P_a, \label{q10d}
\end{align}
\end{subequations}
where constraint \eqref{q10b} denotes the DEP requirement at Willie, $P_k$ and $P_a$ denote the maximum transmission power at the $k$th sensor and the AP, respectively. Because
the optimization variables are coupled in \eqref{q10a} and \eqref{q10b}, Problem \eqref{q10} is non-convex, which is challenging to solve using conventional convex optimization methods.

\section{Exact Penalty Algorithm Based Solution}
In this section, we introduce the exact penalty algorithm to address Problem \eqref{q10} and obtain a locally optimal solution.

We begin by addressing the non-convex constraint \eqref{q10b} and introduce a new variable, denoted as $x\triangleq\frac{\sigma_1}{\sigma_0}$. Consequently, constraint \eqref{q10b} can be reformulated as
\begin{align}
f(x)\triangleq\ln x+\frac{1}{x}\leq 1+2\epsilon^2.
\end{align}

By introducing $x_1$ and $x_2$ as the two roots of $f(x)=1+2\epsilon^2$, we can equivalently transform \eqref{q10b} into
\begin{align}
x_1\leq x \leq x_2,
\end{align}
where
\begin{align}
x_1&=\exp(\mathcal{W}_{-1}(-\exp(-(1+2\epsilon^2)))+1+2\epsilon^2),\\
x_2&=\exp(\mathcal{W}_{0}(-\exp(-(1+2\epsilon^2)))+1+2\epsilon^2),
\end{align}
and $\mathcal{W}(z)$ denotes the Lambert $\mathcal{W}$ function of $z$.
Because $x=1+\frac{\sum_{k=1}^{K} \mathbf{g}_k\mathbf{W}_k\mathbf{W}^H_k\mathbf{g}^H_k}{\mathbf{f}\mathbf{V}\mathbf{V}^H\mathbf{f}^H+\sigma^2_w}>1$, we have
\begin{align}\label{q11}
0<\frac{\sum_{k=1}^{K} \mathbf{g}_k\mathbf{W}_k\mathbf{W}^H_k\mathbf{g}^H_k}{\mathbf{f}\mathbf{V}\mathbf{V}^H\mathbf{f}^H+\sigma^2_w}<x_2-1.
\end{align}

Therefore, Problem \eqref{q10} can be reformulated as
\begin{subequations}\label{q12}
\begin{align}
\min_{\scriptsize\begin{array}{c}\mathbf{W}_k, \mathbf{V}, \mathbf{U}_a
 \end{array} }\ &\mathrm{MSE}_a\label{q12a}\\
\textrm{s.t.}\quad \quad \  & \eqref{q11},\eqref{q10c},\eqref{q10d}.
\end{align}
\end{subequations}

Given $\mathbf{W}_k$ and $\mathbf{V}$, the optimization of $\mathbf{U}_a$ in Problem \eqref{q12} can be transformed into one unconstrained optimization problem, i.e., $\min\limits_{\mathbf{U}_a} \mathrm{MSE}_a$. The optimal solution of $\mathbf{U}_a$ based on the optimal MMSE receiver design principle can be expressed by \cite{YLi22,WFang21,MFu22,CHu22}
\begin{align}\label{q13}
\mathbf{U}^o_a=&(\mathbf{A}+\mathbf{B}+\sigma^2_a\mathbf{I}_{N_r})^{-1}\mathbf{C},
\end{align}
where
\begin{align}
\mathbf{A}=&\sum_{k=1}^K\mathbf{H}_k\mathbf{W}_k\mathbf{W}^H_k\mathbf{H}^H_k,\label{q14}\\
\mathbf{B}=&\mathbf{H}_{aa}\mathbf{V}\mathbf{V}^H\mathbf{H}_{aa}^H,\label{q15}\\
\mathbf{C}=&\sum_{k=1}^K\mathbf{H}_k\mathbf{W}_k.\label{q16}
\end{align}

Moreover, we let
\begin{align}
d=&\sum_{k=1}^K\mathbf{g}_k\mathbf{W}_k\mathbf{W}^H_k\mathbf{g}^H_k,\label{q17}\\
e=&\mathbf{f}\mathbf{V}\mathbf{V}^H\mathbf{f}^H,\label{q18}
\end{align}
the constraint \eqref{q11} can be rewritten as
\begin{align}\label{q19}
d\leq(x_2-1)(e+\sigma^2_w).
\end{align}

Substituting \eqref{q14}-\eqref{q16} into \eqref{q12a}, Problem \eqref{q12} can be rewritten as
\begin{subequations}\label{q20}
\begin{align}
\min_{\scriptsize\begin{array}{c}\mathbf{W}_k, \mathbf{V},\\
\mathbf{A}, \mathbf{B},\\
\mathbf{C}, \mathbf{D}, \mathbf{E}
 \end{array} }\ &KN_s-\mathrm{tr}(\mathbf{C}^H(\mathbf{A}+\mathbf{B}+\sigma^2_a\mathbf{I}_{N_r})^{-1}\mathbf{C})\label{q20a}\\
\textrm{s.t.}\quad  \quad  &\eqref{q10c}, \eqref{q10d}, \eqref{q14}-\eqref{q19}.
\end{align}
\end{subequations}

Because the objective function \eqref{q20a} and the equality constraints, i.e., \eqref{q14}, \eqref{q15}, \eqref{q17}, and \eqref{q18} are non-convex, Problem \eqref{q20} remains non-convex. Thus, the globally optimal solution of the optimization problem \eqref{q20} is challgening to obtain. We can transform \eqref{q14}, \eqref{q15}, \eqref{q17}, and \eqref{q18} into suitable convex forms, enabling us to obtain a locally optimal solution using the constrained concave-convex procedure (CCCP)-based method.

\emph{Lemma 1}: Assuming $\mathbf{Y}\succ\mathbf{0}$, $\mathbf{\Omega}=\mathbf{X}^H\mathbf{Y}^{-1}\mathbf{X}$ is equivalent to
\begin{equation}\label{q21a}
   \left[ \begin{array}{ccc}
\mathbf{\Omega} & \mathbf{X}^H \\
\mathbf{X} & \mathbf{Y}\\
\end{array}\right]\succeq \mathbf{0},
\end{equation}
and
\begin{align}\label{q21b}
\mathrm{tr}(\mathbf{\Omega}-\mathbf{X}^H\mathbf{Y}^{-1}\mathbf{X})\leq0.
\end{align}

\emph{Proof}: See Appendix A. $\hfill\blacksquare$

Let us define the auxiliary variables $\mathbf{T}$ and $\mathbf{s}$ as
\begin{align}
\label{qt1}\mathbf{T}=&[\mathbf{H}_1\mathbf{W}_1, \cdot\cdot\cdot, \mathbf{H}_K\mathbf{W}_K],\\
\label{qt2}\mathbf{s}=&[\mathbf{g}_1\mathbf{W}_1, \cdot\cdot\cdot, \mathbf{g}_K\mathbf{W}_K].
\end{align}
By leveraging Lemma 1, the equality constraints, i.e., \eqref{q14}, \eqref{q15}, \eqref{q17}, and \eqref{q18} can be equivalently expressed as
\begin{equation}\label{q22}
   \left[ \begin{array}{ccc}
\mathbf{A} & \mathbf{T} \\
\mathbf{T}^H & \mathbf{I}\\
\end{array}\right]\succeq \mathbf{0},
\end{equation}
\begin{align}\label{q23}
\mathrm{tr}(\mathbf{A}_a-\mathbf{T}\mathbf{T}^H)\leq0,
\end{align}
\begin{equation}\label{q26}
   \left[ \begin{array}{ccc}
\mathbf{B} & \mathbf{H}_{aa}\mathbf{V} \\
\mathbf{V}^H\mathbf{H}_{aa}^H & \mathbf{I}\\
\end{array}\right]\succeq \mathbf{0},
\end{equation}
\begin{align}\label{q27}
\mathrm{tr}(\mathbf{B}-\mathbf{H}_{aa}\mathbf{V}\mathbf{V}^H\mathbf{H}_{aa}^H)\leq0,
\end{align}
\begin{equation}\label{q28}
   \left[ \begin{array}{ccc}
d & \mathbf{s} \\
\mathbf{s}^H & \mathbf{I}\\
\end{array}\right]\succeq \mathbf{0},
\end{equation}
\begin{align}\label{q29}
\mathrm{tr}(d-\mathbf{s}\mathbf{s}^H)\leq0.
\end{align}
\begin{equation}\label{q30}
   \left[ \begin{array}{ccc}
e & \mathbf{f}\mathbf{V} \\
\mathbf{V}^H\mathbf{f}^H & \mathbf{I}\\
\end{array}\right]\succeq \mathbf{0},
\end{equation}
\begin{align}\label{q31}
\mathrm{tr}(e-\mathbf{f}\mathbf{V}\mathbf{V}^H\mathbf{f}^H)\leq0.
\end{align}

Therefore, Problem \eqref{q20} can be recast as
\begin{subequations}\label{q32}
\begin{align}
\min_{\Xi}\ &KN_s-\mathrm{tr}(\mathbf{C}^H(\mathbf{A}+\mathbf{B}+\sigma^2_a\mathbf{I}_{N_r})^{-1}\mathbf{C})\label{q30a}\\
\textrm{s.t.}\ &\eqref{q10c}, \eqref{q10d}, \eqref{q16}, \eqref{q19}, \eqref{qt1}-\eqref{q31},
\end{align}
\end{subequations}
where $\Xi=\{\mathbf{W}_k, \mathbf{V}, \mathbf{T}, \mathbf{s}, \mathbf{A}, \mathbf{B}, \mathbf{C}, d, e\}$.
In Problem \eqref{q32}, the constraints \eqref{q22}, \eqref{q26}, \eqref{q28}, and \eqref{q30} are linear matrix inequalities (LMIs). Constraints \eqref{q10c} and \eqref{q10d} are convex quadratic. Constraints \eqref{q16}, \eqref{q19}, \eqref{qt1}, and \eqref{qt2} are linear. However, the objective function \eqref{q30a}, constraints \eqref{q23}, \eqref{q27}, \eqref{q29}, and \eqref{q31} remain nonconvex and they possess the DC form. To solve Problem \eqref{q32}, we first transform Problem \eqref{q32} to a DC programming problem, and then obtain a locally optimal solution by using the penalty based algorithm.

Consequently, let us define $\Upsilon\triangleq\{\mathbf{A}, \mathbf{B}, \mathbf{C}\}$, $\Gamma\triangleq\{\mathbf{A}, \mathbf{B}, d, e\}$, $\Psi\triangleq\{\mathbf{T}, \mathbf{s}, \mathbf{V}\}$, and the functions
\begin{align}\label{q33}
f(\Upsilon)=&\mathrm{tr}(\mathbf{C}^H(\mathbf{A}+\mathbf{B}+\sigma^2_a\mathbf{I}_{N_r})^{-1}\mathbf{C}),\\
f(\Gamma)=&\mathrm{tr}(\mathbf{A}+\mathbf{B}+d+e),\\
f(\Psi)=&\mathrm{tr}(\mathbf{T}\mathbf{T}^H+\mathbf{H}_{aa}\mathbf{V}\mathbf{V}^H\mathbf{H}_{aa}^H+\mathbf{s}\mathbf{s}^H+\mathbf{f}\mathbf{V}\mathbf{V}^H\mathbf{f}^H).
\end{align}

Employing the exact penalty based algorithm \cite{QLi18,URashid14}, we can rewrite Problem \eqref{q32} as
\begin{align}\label{q36}
\min_{\Xi\in\Theta}\ &KN_s-f(\Upsilon)+p(f(\Gamma)-f(\Psi))
\end{align}
where
\begin{align}\label{q37}
\Theta\triangleq\{\Xi|&\eqref{q10c}, \eqref{q10d}, \eqref{q16}, \eqref{q19}, \eqref{qt1}, \eqref{qt2}, \eqref{q22}, \eqref{q26}, \eqref{q28}, \eqref{q30}\}
\end{align}
is a compact set, and $p$ denotes a penalty factor.

The equivalency between Problem \eqref{q32} and Problem \eqref{q36} is provided by the following lemma, which is proved in \cite{QLi18,URashid14}.

\emph{Lemma 2}: There is a limited $+\infty>p_0>0$ such that when $p>p_0$, Problem \eqref{q36} is equivalent to Problem \eqref{q32}. $\hfill\blacksquare$

According to Lemma 2, we can solve Problem \eqref{q36} by the CCCP-based algorithm. The first-order Taylor expansions of $f(\Upsilon)$ and $f(\Psi)$ around the point $\bar{\Upsilon}$ and $\bar{\Psi}$ are calculated as
\begin{align}
f(\Upsilon;\bar{\Upsilon})= &-\mathrm{tr}(\bar{\mathbf{C}}^H(\bar{\mathbf{A}}+\bar{\mathbf{B}}+\sigma^2_a\mathbf{I}_{N_r})^{-1}\bar{\mathbf{C}})\nonumber\\
&+2\mathbb{R}\{\mathrm{tr}(\bar{\mathbf{C}}^H(\bar{\mathbf{A}}+\bar{\mathbf{B}}+\sigma^2_a\mathbf{I}_{N_r})^{-1}\mathbf{C})\}\nonumber\\
&-\mathbb{R}\{\mathrm{tr}(\bar{\mathbf{C}}^H(\bar{\mathbf{A}}+\bar{\mathbf{B}}+\sigma^2_a\mathbf{I}_{N_r})^{-1}(\mathbf{A}+\mathbf{B}-\bar{\mathbf{A}}-\bar{\mathbf{B}})\nonumber\\
&\cdot(\bar{\mathbf{A}}+\bar{\mathbf{B}}+\sigma^2_a\mathbf{I}_{N_r})^{-1}\bar{\mathbf{C}})\},\\
f(\Psi;\bar{\Psi})=&-\mathrm{tr}(\bar{\mathbf{T}}\bar{\mathbf{T}}^H+\mathbf{H}_{aa}\bar{\mathbf{V}}\bar{\mathbf{V}}^H\mathbf{H}_{aa}^H+\bar{\mathbf{s}}\bar{\mathbf{s}}^H+\mathbf{f}\bar{\mathbf{V}}\bar{\mathbf{V}}^H\mathbf{f}^H)\nonumber\\
&+2\mathbb{R}\{\mathrm{tr}(\mathbf{T}\bar{\mathbf{T}}^H+\mathbf{H}_{aa}\mathbf{V}\bar{\mathbf{V}}^H\mathbf{H}_{aa}^H+\mathbf{s}\bar{\mathbf{s}}^H\nonumber\\
&+\mathbf{f}\mathbf{V}\bar{\mathbf{V}}^H\mathbf{f}^H)\}.
\end{align}

Assume $\left(\Upsilon^{(m)}, \Psi^{(m)}\right)$ is optimal in the $m$th iteration, Problem \eqref{q36} can be expressed as
\begin{align}\label{q39}
\min_{\Xi\in\Theta}\ &KN_s-f(\Upsilon;\Upsilon^{(m)})+p(f(\Gamma)-f(\Psi;\Psi^{(m)})).
\end{align}
in the $(m+1)$th iteration. The aforementioned procedures for solving Problem \eqref{q11} are summarized in Algorithm 1, where the optimal $\mathbf{A}, \mathbf{B}$, $\mathbf{C}, d, e$, $\mathbf{s}$, $\mathbf{W}_k$, $\mathbf{T}$, and $\mathbf{V}$ in the $m$th iteration are denoted as $\mathbf{A}^{(m)}, \mathbf{B}^{(m)}$, $\mathbf{C}^{(m)}, d^{(m)}, e^{(m)}$, $\mathbf{s}^{(m)}$, $\mathbf{W}^{(m)}_k$, $\mathbf{T}^{(m)}$, and $\mathbf{V}^{(m)}$, respectively.

\emph{Complexity Analysis}: Problem \eqref{q39} can be solved by the interior-point method \cite{IPolik10}. Thus, the computational complexity of Algorithm 1 is $\mathcal{O}(L_1(KN_s^2+N_t^2+N_rKN_s+KN_s+2N_r^2+NrN_s+2)^{3.5}\log \frac{1}{\epsilon})$, where $L_1$ is the number of iterations for the convergence of Algorithm 1 and $\epsilon$ is the accuracy.

\begin{algorithm}[h]
\caption{Proposed Exact Penalty Based Algorithm}
\begin{algorithmic}[1]
\STATE \textbf{Initialize:} $m=0$, $p$, $\mathbf{A}^{(0)}, \mathbf{B}^{(0)}, \mathbf{C}^{(0)}$, $\mathbf{T}^{(0)}$, $\mathbf{s}^{(0)}$, and $\mathbf{V}^{(0)}$;
\STATE \textbf{Repeat} \\
\quad $m:=m+1$;         \\
\quad Update $\mathbf{W}^{(m)}_k$, $\mathbf{T}^{(m)}$, $\mathbf{V}^{(m)}$, $\mathbf{A}^{(m)}$, $\mathbf{B}^{(m)}$,  $\mathbf{C}^{(m)}$, $\mathbf{s}^{(m)}$, $d^{(m)}$, and $e^{(m)}$ by solving Problem \eqref{q39};\\
\STATE \textbf{Until:} Convergence.
\end{algorithmic}
\end{algorithm}

\section{Numerical Results}
In simulations, we assume that all the channel responses are independent and identically distributed
complex Gaussian random variables with zero mean and unit variance \cite{CHu22}. We consider a scenario where $N_s=4$, $N_t=4$, and $N_r=4$. The noise powers set $\sigma_w^2=\sigma_a^2=\sigma^2$.  We also assume that all sensors have an identical maximum transmission power, leading to a consistent SNR throughout the simulation, given by $P_{1}/\sigma^2=\cdots=P_{K}/\sigma^2=P_s/\sigma^2$ \cite{CHu22}.  The maximum transmitted SNR at the AP is set to $P_a/\sigma^2=30$ dB.  To evaluate the performance, we employ the normalized MSE, denoted as $\mathrm{MSE}_a/K$ \cite{XLi19}.  Additionally, we use a self-interference coefficient of $\rho=0.5$ and a tolerated detection coefficient of $\epsilon=0.1$.

Fig. 1 illustrates the convergence behavior of our proposed exact penalty-based algorithm for different values of $P_s/\sigma^2$ ($5$ dB, $10$ dB, and $25$ dB). The results in Fig. 1 indicate that the algorithm converges after approximately 50 iterations.

\begin{figure}
\centering
\includegraphics[width=2.4in]{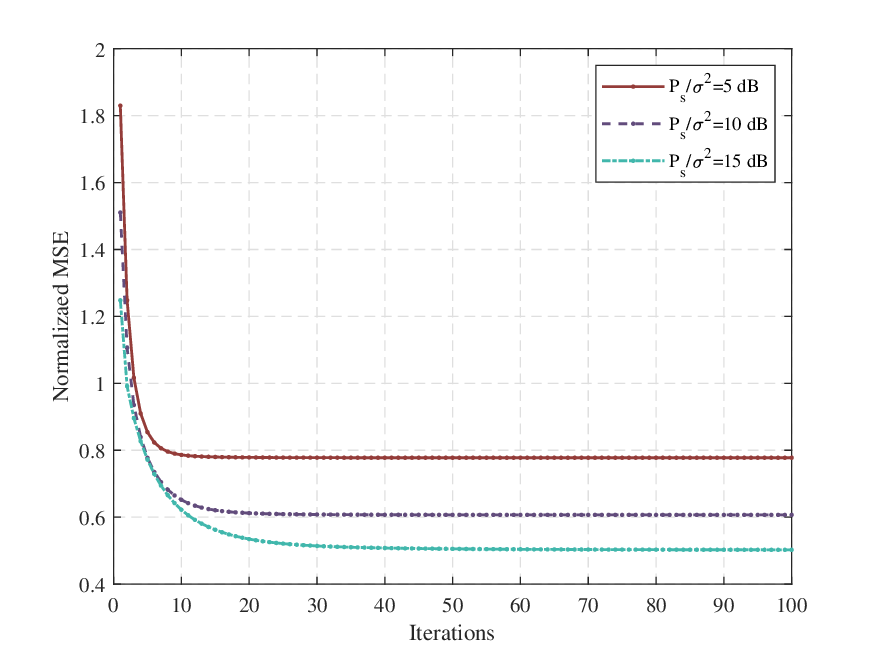}
\caption{The normalized MSE versus the number of iterations; convergence behavior of our proposed exact penalty based algorithm, where $P_s/\sigma^2=5$ dB, 10 dB, 15 dB, $K=10$.}
\end{figure}

Fig. 2 illustrates the impact of $P_s/\sigma^2$ on the normalized MSE with a total of $K=10$ sensors. In the legend, ``Proposed" represents our proposed scheme; ``Random AN" stands for the random artificial noise (AN) scheme; ``MRT AN" corresponds to the maximal ratio transmission (MRT) scheme, and ``w/o AN" denotes our proposed scheme without AN. It can be observed from Fig. 2 that the ``Proposed" scheme consistently outperforms the other schemes in terms of MSE performance. As $P_s/\sigma^2$ increases, the normalized MSE decreases for all schemes. Furthermore, the gaps in normalized MSE between the ``Proposed" scheme and the other schemes, except for the ``w/o AN" scheme, decrease with increasing $P_s/\sigma^2$. This phenomenon can be attributed to the fact that the DEP requirement of Willie becomes easier to satisfy with the assistance of AN, resulting in a smaller MSE at the AP. Beyond a $P_s/\sigma^2$ value of $15$ dB, the performance of the ``w/o AN" scheme levels off, indicating that increasing the maximum transmission power of sensors no longer leads to performance improvements in this scheme.
\begin{figure}
\centering
\includegraphics[width=2.4in]{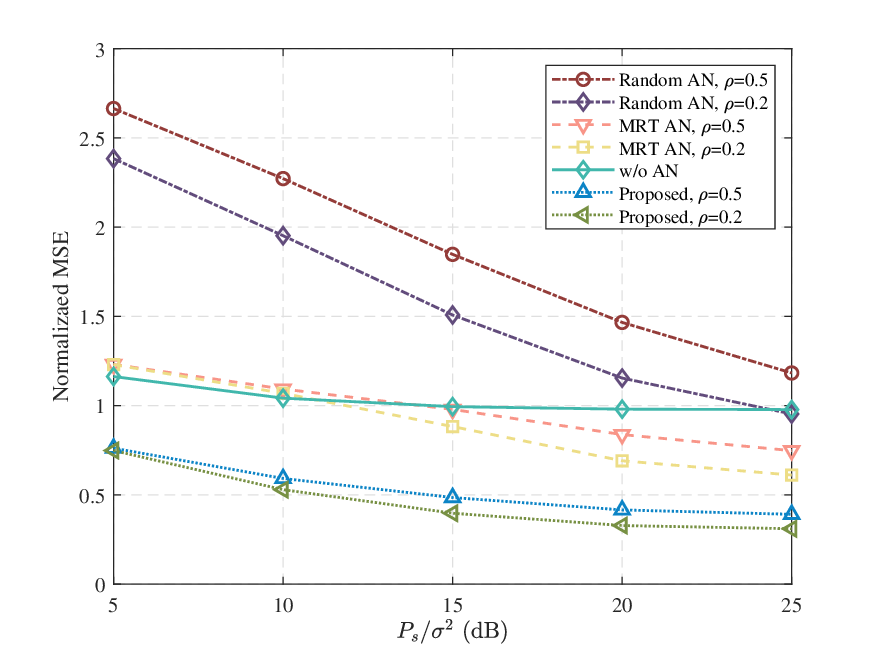}
\caption{The normalized MSE versus  $P_s/\sigma^2$; performance comparison of ``Proposed", ``Random AN", ``MRT AN", and ``w/o AN" schemes, where $K=10$.}
\end{figure}

Fig. 3 investigates the influence of the number of sensors $K$ on the normalized MSE, with a fixed $P_s/\sigma^2$ value of $10$ dB. As shown in Fig. 3, the ``Proposed" scheme consistently outperforms the other schemes in terms of MSE performance. Notably, the normalized MSE decreases for all schemes as $K$ increases. Furthermore, the gaps in normalized MSE between the ``Proposed" scheme and the other schemes decrease with increasing $K$. This is due to the performance gain introduced by our proposed scheme becoming distributed more thinly across the growing number of sensors, resulting in smaller individual improvements on average.

\begin{figure}
\centering
\includegraphics[width=2.3in]{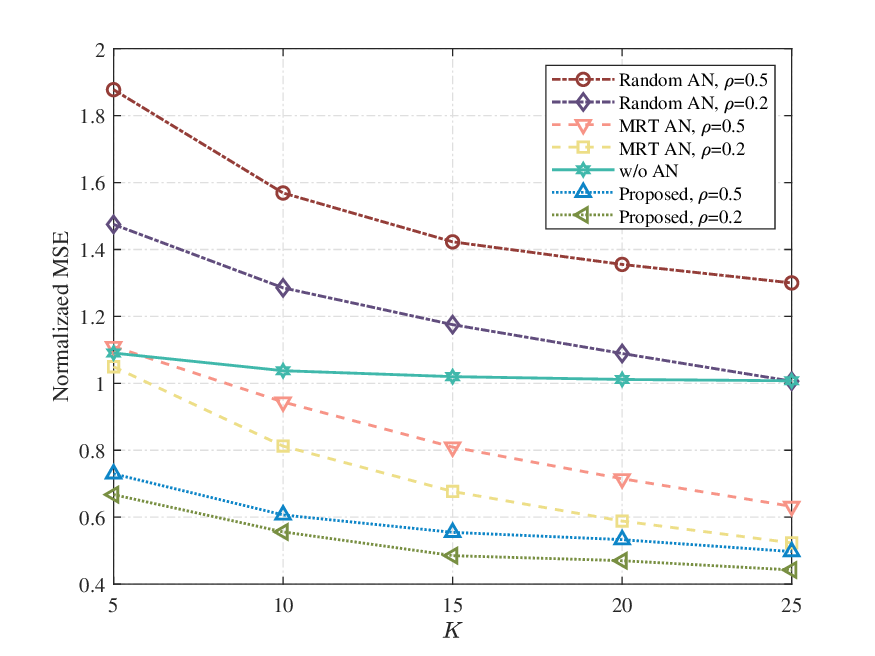}
\caption{The normalized MSE versus the number of sensors $K$; performance comparison of ``Proposed", ``Random AN", ``MRT AN", and ``w/o AN" schemes, where $P_s/\sigma^2=10$ dB.}
\end{figure}

In Fig. 4, we explore the influence of the tolerated detection coefficient $\epsilon$ on the normalized MSE, with $P_s/\sigma^2=10$ dB and $K=10$. It is evident from Fig. 4 that the ``Proposed" scheme outperforms the other schemes in terms of MSE performance.  As $\epsilon$ increases, both the ``Proposed" and ``w/o AN" schemes exhibit decreasing normalized MSEs, and the gap between them narrows. However, the curves for the ``Random AN" and ``MRT AN" schemes remain relatively flat, as the DEP requirement of Willie does not yield further performance improvement in these cases.
\begin{figure}
\centering
\includegraphics[width=2.3in]{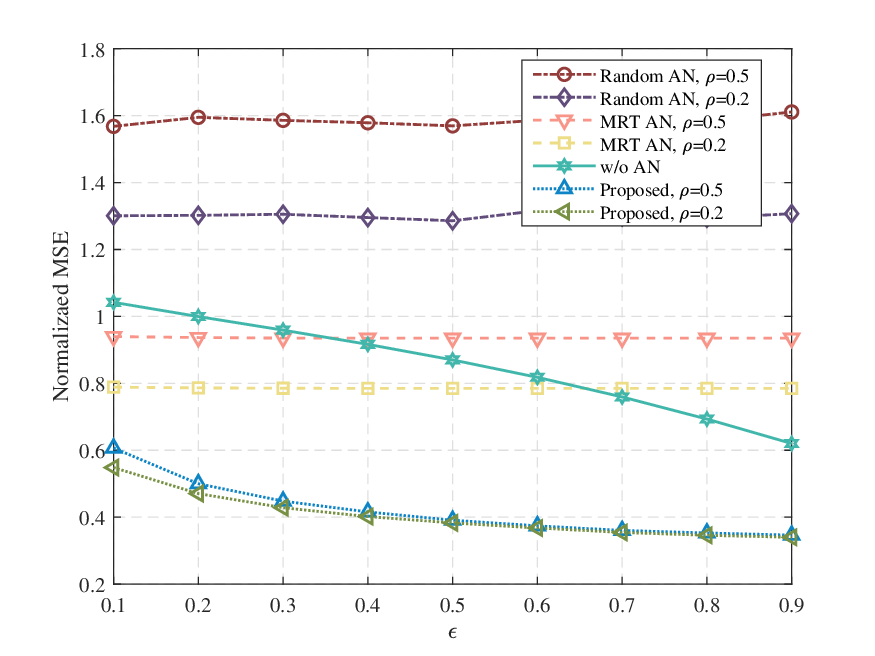}
\caption{The normalized MSE versus the tolerated detection coefficient $\epsilon$; performance comparison of ``Proposed", ``Random AN", ``MRT AN", and ``w/o AN" schemes, where $P_s/\sigma^2=10$ dB and $K=10$.}
\end{figure}

\section{Conclusion}

In this paper, we considered the beamforming optimization in an MIMO AirComp system under the convertness constraint, where the AP sends the AN to improve covert performance. We considered the MSE of the AP minimization problem under the power constraints of the AP and the sensors,  while also considering the DEP requirement of Willie. The optimization problem was non-convex due to the coupled variables. Thus, we proposed an exact penalty based algorithm to solve it and obtained the locally optimal solution. Numerical simulations underscored the superior MSE performance achieved by our proposed optimization algorithm compared to benchmark schemes.

\appendices
\section{Proof of Lemma 1}
Upon using the Schur complement, \eqref{q21a} is equivalent to
\begin{align}\label{aq1}
\mathbf{\Omega}-\mathbf{X}^H\mathbf{Y}^{-1}\mathbf{X}\succeq \mathbf{0}.
\end{align}
Upon combining \eqref{aq1} with \eqref{q21b}, we have
\begin{align}\label{aq2}
\mathrm{tr}(\mathbf{\Omega}-\mathbf{X}^H\mathbf{Y}^{-1}\mathbf{X})=0.
\end{align}
From \eqref{aq1} and \eqref{aq2}, we obtain
\begin{align}\label{aq3}
\mathbf{\Omega}=\mathbf{X}^H\mathbf{Y}^{-1}\mathbf{X}.
\end{align}

\end{document}